\documentstyle[12pt,aasms4]{article}
\def\gray{{$\gamma$-ray\ }}
\def\grays{{$\gamma$-rays\ }}
\def\etal{{\it et al.\ }}
\def\mic{{$\mu$m\ }}
\def\eg{{\it e.g.}}
\begin{document}

\title{AN EMPIRICALLY BASED CALCULATION OF THE EXTRAGALACTIC INFRARED
       BACKGROUND}

\author{M. A. Malkan}
\affil{Department of Physics and Astronomy, University of California,
Los Angeles,\\ Los Angeles, CA 90095-1562}
\authoremail{malkan@bonnie.astro.ucla.edu}

\author{F. W. Stecker}
\affil{Laboratory for High Energy Astrophysics, NASA Goddard Space
Flight Center,\\ Greenbelt, MD 20771}
\authoremail{stecker@lheavx.gsfc.nasa.gov}

\author{\it The Astrophysical Journal, in press.}

\begin{abstract}

Using the excellent observed correlations among 
various infrared wavebands with
12 and 60\mic luminosities, we calculate the 2-300\mic
spectra of galaxies as a function of luminosity.
We then use 12 \mic and 60 \mic galaxy luminosity functions 
derived from IRAS data, 
together with recent data on the redshift evolution of galaxy emissivity,
to
derive a new, empirically based IR background spectrum from stellar and
dust emission in galaxies.
Our best estimate for the IR background is of order
2-3 nW m$^{-2}$ sr$^{-1}$
with a peak around 200\mic reaching $\sim$ 6-8 nW m$^{-2}$
sr$^{-1}
$. Our empirically derived background spectrum is fairly flat in the mid
IR, as opposed to spectra based on modeling with discrete temperatures which
exhibit a pronounced ``valley'' in the mid-IR.
We also derive a conservative lower limit to the IR background which
is more than a factor of 2 lower than our derived flux.

\end{abstract}


\section{Introduction}

The extragalactic diffuse IR background has long been recognized
to contain important information about the evolution
of galaxies. More recently, studies of the IR background have become
important to very-high-energy \gray astronomers because pair-production
interactions of multi-TeV \grays with extragalactic IR photons are an important
source of extinction of extragalactic \grays (Stecker \& De Jager
1977a and references therein). Indeed, one empirical way to determine the
IR background flux is to look for absorption of TeV \grays in the spectra
of blazars (Stecker, De Jager \& Salamon 1992).

Stecker, Puget \& Fazio (1977) made an early estimate of the extragalactic IR
background and showed that if galaxies radiate a significant fraction of
their energy from stellar nucleosynthesis in the mid to far IR range, the
background flux would be of the order of 10 nW m$^{-2}$sr$^{-1}$. The 
COBE-DIRBE detector has now obtained extragalactic residual flux measurements 
which approach or reach this range (Hauser 1996). There have been many 
attempts to
model the expected IR background from galaxies; they have recently been
reviewed by Lonsdale (1996).

In this letter, we take an alternate approach. Rather than
modeling the IR emission from galaxies using assumptions such as theoretical
dust temperature distributions, we base our calculations on empirical
studies of the IR spectra of galaxies with a wide distribution of
intrinsic luminosities, and on empirically determined luminosity
functions for these galaxies.

\section{Infrared Spectra of Galaxies }

The starlight spectra of all but the youngest galaxies have a peak
in $\nu L_\nu$ from red giants around 1--1.5\mic,
and drop as a Jeans tail at wavelengths longer than 2\mic.
At wavelengths longer than 3\mic, spiral galaxies emit by thermal
radiation from dust grains, which are warmed to a very wide range of
temperatures, from a minimum of 15--20 K to a maximum which can exceed
800 K (Joseph, Meikle, Robertson \& Wright 1984).
Because of this very large temperature range, 
modelling of the diffuse mod-IR and far-IR background
from thermal dust emission from a large sample of galaxies  
by choosing one, or even two temperatures is not
realistic. This is why our empirical appoach is preferable to what
has been done previously.

Observations with the Infrared Astronomy Satellite
(IRAS), combined with
other data, have indicated that the infrared (10$^{12}$
to 10$^{14}$ Hz) spectra of
all types of spiral galaxies depend systematically on their total
luminosity.
There are well-determined relations between the luminosity emitted
in a given infrared waveband with the luminosity at 60$\mu$m.
Spinoglio, \etal(1995) established tight empirical correlations
relating 100, 60, 25, 12\mic and 1.2--2.2\mic (NIR) luminosities
with the bolometric luminosity of Seyfert and non-Seyfert galaxies.
We have
combined these relations so that for any given value of $L_{12}$
or  $L_{60}$,
we can predict the luminosities at all other infrared wavelengths.
The luminosities at 2.2 and 3.5\mic in $W Hz^{-1}$ were estimated 
from the relation for
$L_{NIR}$ (given in Watts), as $L_{3.5} = 1.50 L_{2.2}$  
= 2.05 $\times 10^{-14}L_{NIR}$
for Seyfert 1 galaxies,
$L_{3.5} = 1.08 L_{2.2} = 1.20 \times 10^{-14} L_{NIR}$ for Seyfert 2 galaxies,
and $L_{3.5} = 1.06 L_{2.2} = 1.01 \times 10^{-14} L_{NIR}$ 
for  non-Seyfert galaxies.
We have extrapolated to wavelengths longer than 100\mic using the
correlation of far-infrared color temperature with the parameter
$\alpha _{60-100}$
given by Spinoglio, \etal(1995) in their Appendix B,
derived from far-infrared photometry data obtained from the Kuiper Airborne 
Observatory.
Luminosities at all other wavelengths were then estimated by linear
interpolation in $Log(\nu)$ versus $Log(L_{\nu})$.

The largest proportions of hot dust are found in galaxies with the
highest rates of current star formation.  This results in a systematic
trend for more luminous galaxies to emit relatively hotter infrared
spectra.  We illustrate this gradual trend in Figure 1, which shows
the IR spectra of galaxies with luminosities taken differing by
various factors of 10 from $L_{*}$ = 10$^{22.2}$ W Hz$^{-1}$, 
ranging from 10$^{-3}$ $L_{*}$ to 10$^{3}$ $L_{*}$.
The luminosity correlations among the various IR wavebands are very
tight for all kinds of non-Seyfert spiral galaxies ($\sim$ 90\% of
spirals),
including those classified as ``normal", ``LINER", and ``Ultraluminous".
Thus over more than four orders of magnitude of luminosity, the infrared
spectra of galaxies change gradually and continuously.
The data do not support any clear quantitative distinction into
separate categories of galaxies (such as ``starbursts" or
``hyperluminous" or ``dwarfs"). For this reason, we can use a single
luminosity
function (LF) obtained from 60\mic surveys to represent all IR emitting
galaxies. (Elliptical galaxies, being dust-poor, are not significant
enough mid-IR and
far-IR emitters to be included in our calculation).

\section{Galaxy Luminosity Functions }

The total diffuse IR background at a given frequency is calculated by
integrating over the luminosity function of galaxies
and then integrating over redshift. 
(See, \eg, the formalism in Stecker \& Salamon (1996) for a derivation 
of diffuse background flux from unresolved point sources).
We have used two different assumed galaxy luminosity functions,
both of which are derived directly from IRAS source catalogs having
complete spectroscopic redshift measurements.
The first is the Extended 12 Micron Galaxy Sample (Rush, Malkan \&
Spinoglio 1993), which is based on a 12\mic flux-limited sample of
893 galaxies.
The second is the sample of 2818 galaxies selected at 60\mic by Saunders 
\etal 1990).
Both differential luminosity functions are adequately fitted by a double-power-law
function of the form:
\begin{equation}
\label{LF}
\phi = C (L/L_{*})^{1-\alpha}[1 + (L/\beta L_{*})]^{-\beta}
\end{equation}
with low-luminosity slopes of $\alpha = 1.35 $ and 1.7 for the respective
12\mic and 60\mic LF's,
steepening by $\beta = 2.11 $ and 3.6 respectively, around a ``knee"
at $L_{*} = 10^{24.22}h_{50}^{-2}$W Hz$^{-1}$ 
and $ 10^{23.0}h_{50}^{-2}$W Hz$^{-1}$, respectively. 
The value given by Saunders, \etal (1990) for the normalization constant 
of the 60\mic LF is $C_{60} = 3.25 \times 10^{-3}h_{50}^{3}$ 
Mpc$^{-3}$dex$^{-1}$.
The value for the 12\mic normalization constant, obtained 
from the data of Rush, \etal (1993), is $C_{12}  = 2.7 
\times 10^{-3}h_{50}^{3}$ Mpc$^{-3}$dex$^{-1}$.

The Saunders, \etal (1990) luminosity function is similar to that of the
QDOT survey (Lawrence, \etal 1986) and the BGS (Soifer, \etal 1987).
In particular all of these LF's agree extremely well on the volume density
of galaxies within an order-of-magnitude of $L_{*}$.  We confirmed that
our results are hardly changed (by less than 10\% at any frequency) by
substituting any of these 60\mic LFs into our integration.

\section{The Background Calculation}

In order to calculate the total contribution to the IR background from
sources emitting at various redshifts, we must extrapolate 
the luminosity function data, obtained mainly for 
low-redshift galaxies, all the way to the redshift
of initial galaxy formation, $z_{max}$.
At present, IR observations of galaxies at high redshifts are severely limited.
Although optical and ultraviolet searches tend to discover high-redshift
galaxies with apparent visual absorptions of less than half a magnitude,
near-infrared searches are capable of finding substantially dustier
galaxies (Malkan, Teplitz and McLean 1995; 1996).
A recent deep ISOCAM 7\mic image of the Hubble Deep Field
appears to have reliably detected 13 galaxies at $z \le 0.5$ 
(Rowan-Robinson, \etal 1997).
Assuming these identifications are correct, they imply a very
high rate of mid-infrared emission from luminous IR galaxies at z=1.

The redshift evolution of the infrared galaxy LF is only weakly
constrained at present.
Saunders, \etal (1990) found that their data could be fitted by a luminosity
evolution with
Q = $3 \pm 1$. A more recent analysis of a deep sub-sample of 
the IRAS Faint Source Survey by Bertin, Dennefeld \& Moshir (1997) finds
evidence for strong evolution of IRAS galaxies with $Q = 3.2\pm 0.2 \pm 0.3$
out to a redshift of 0.3. Previous studies by Soifer, \etal (1987), Lonsdale,
\etal (1990) and Pearson \& Rowan-Robinson (1996) have come to similar 
conclusions regarding the redshift evolution of IRAS sources. The analysis
of Gregorich, \etal (1995), which indicated even stronger evolution, and 
which would lead to predictions of a higher IR background, appears to
disagree with other analyses; their sample may have been contaminated by
false detections (Bertin, \etal 1997).

Lilly, \etal (1996) found strong galaxy evolution at 280 nm, a wavelength which
reflects instantaneous (rather than cumulative) star formation activity. Their
value for Q was was in the range of 3 to 4, depending on
the cosmological model chosen.
If, as is
generally believed, short-wavelength emission from hot, young O and B
stars is reradiated by dust in the far IR, a consistency with the IRAS
and ISO studies emerges.
Pei \& Fall (1995) analysed the evolution of neutral hydrogen gas in damped
Ly$\alpha$ systems seen in quasar spectra.
Their study indicated that the star formation rate in the universe
peaks at a redshift between 1 and 2, evolving roughly 
as $(1 + z)^{3\pm 1}$
at lower redshifts and falling off at higher redshifts.
Such evolution is consistent with
recent studies of the redshift distribution
of galaxy emissivity and corresponding star formation history 
by Lilly, \etal (1996), Cowie, Songaila, Hu \& Cohen (1996), 
Madau, \etal (1996), and Connelley, \etal (1997).

We have therefore evolved the locally-determined galaxy LF to higher redshift
assuming (1) pure luminosity evolution of the form
$L(z) = L_{0}(1 + z)^Q$ for $z \le z_{flat}$, (2) no evolution, {\it i.e.},
$L(z) = L_{0}(1 + z_{flat})^Q$ = const, for $z_{flat} \ge z \ge z_{max}$,
and (3) no emission, {\it i.e.}, $L(z) = 0$, for $z \ge z_{max}$. 
Since it has been fairly well established that the stellar emissivity from
galaxies peaks at a redshift between 1 and 2 (see above), we assumed values
for $z_{flat}$ of 1 and 2, $z_{max}$ = 4, and Q = 3 
for our ``best upper and lower estimates''.
(We checked that pure density evolution $\propto (1 + z)^7$, 
which has less physical plausibility, gave very similar results.)

The upper two (heavier) solid curves in Figure 2 show 
the result of our computation with
Q = 3 for values of $z_{flat}$ of 1 and 2 and a $z_{max}$ of 4.
(The solid lines are based on the 60\mic LF, while the dashed lines
were obtained using the 12\mic LF with only the $z_{flat}=2$ case and the 
conservative lower limit shown.)  The estimated IR background is on average
$\sim$20\% higher using the 60\mic LF.
(Note that, as given in the 
discussion following Eq. (1), $C_{60}$ is greater than $C_{12}$).
This might be attributed partly to some possible incompleteness in the 12\mic 
survey versus the 60\mic surveys with better sensitivity. We also note that 
the 12\mic LF refers only to non-Seyfert galaxies, and therefore 
has a very steep high-luminosity cut-off.
The 60\mic LF, which does not separate out the Seyferts, 
therefore includes some extremely red galaxies with peak luminosities
around 60\mic that were not included in the 12\mic calculation.

Seyfert galaxies make up approximately 10\% of the IR background at 12\mic
and even less at longer wavelengths where non-Seyfert galaxies are 
relatively brighter.
A crude indication of the relative contributions to the 12\mic
background from low-redshift
Seyferts can be inferred from the fact that 53 of the 893 (6\%)
12 Micron Galaxies are classified as type 1, 
and 63 (7\%) are classified as type 2.
We have calculated their contribution to the diffuse background and
determined that it is less 
than 10\%, 
assuming the same evolution as non-Seyferts.

A very conservative lower limit on the IR background spectrum
was derived by assuming the least amount of galaxy luminosity evolution
consistent with the IRAS galaxy counts,
{\it i.e.}, Q=2 and a maximum redshift $z_{max} = 1$.
This is the smallest $z_{max}$ and evolution exponent which are allowed 
(\eg, Rowan-Robinson \etal, 1997).
The conservative lower limit is shown by the lower light lines in Figure 2
(with the 60\mic LF calculation as the solid line and the 12\mic
calculation as the dotted line).
The amplitude of this lower limit flux is less than half of that calculated 
for the $z_{flat}$ = 1, $z_{max}$ = 4 estimate 
and less than one third of that 
calculated for the $z_{flat}$ = 2 $z_{max}$ = 4 estimate.
The predicted shape of the lower-limit background spectrum is similar 
to that of our $z_{flat}$ = 1 estimate; 
it differs from the $z_{flat}$ = 2 case because the peak is
shifted to slightly shorter wavelengths owing to the redshift cut-off.
For similar reasons, the 3\mic bump from redshifted starlight is relatively
weaker, since we have cut off emission beyond z = 1.

\section{Discussion and Conclusions}

We can compare our best-estimate IR background spectra (the darker
solid lines in Figure 2) with the
presently existing data, and with previous calculations based on modelling.
Figure 2 shows the COBE-DIRBE residuals given by Hauser (1996),
upper limits determined from a fluctuation analysis of the COBE-DIRBE
maps by Kashlinsky, \etal (1996), and the upper limit obtained from the lack
of significant absorption in the \gray spectrum of Mrk 421 up to $\sim$
3 TeV by Stecker and De Jager (1993). Our results are also consistent
with upper limits obtained by Dwek \& Slavin (1994), also using TeV \gray
data from Mrk 421. A power-law extrapolation of our results 
to wavelengths greater than 500\mic would be consistent 
with upper limits obtained in that range
from the FIRAS data by Fixsen, \etal (1996). Our best-estimate background
spectrum agrees quite well with the
claimed detection levels at wavelengths greater than 150\mic obtained by 
Puget, \etal (1996) (not shown) based on an analysis of FIRAS data.

There are several uncertainties which could affect our results:

(1) The redshift corresponding to the maximum star formation
rate is, at present, not well defined.

(2) Uncertainties in the galaxy LF at low luminosities do not affect our result
much because the integrated infrared luminosity is dominated by the 
contributions of galaxies within an order of magnitude of the luminosity
knee at $L_{*}$.

(3) The contributions from galaxies at redshifts 
greater than those where the luminosity function is strongly evolving
are substantially reduced by the cosmological factors involving geometry
and energy redshifting.
For this reason, our choice of the maximum redshift does not influence the 
results much. There is a more significant dependence on the redshift, 
$z_{flat}$, below which we assume $(1+z)^3$ luminosity evolution and beyond 
which we assume that no evolution occurred.

Our results are within 50\% of previous results obtained by theoretical
modeling (see review by Lonsdale 1996). They are in general agreement
in flux level with the most recent calculation of Franceschini, \etal (1997).
However, our spectrum, being based on empirical data, is somewhat 
different from many previous results 
which are based on theoretical modeling, {\it e.g.}, Franceschini \etal. (1997)
and previous work).
In particular, the ``valley'' between stellar emission at a few \mic and dust
emission at $\sim$ 100\mic is almost non-existent in our spectrum,
because our empirical galaxy spectra automatically account for the
very wide range of dust temperatures present. 

The extragalactic infrared background fluxes calculated in this paper have been
used to make new calculations of the absorption of extragalactic high
energy \grays. The results of those calculations will be presented elsewhere 
(Stecker \& De Jager 1997b,c).

\clearpage
\section*{Figure Captions}

Figure 1. Observed dependence of infrared spectra of non-Seyfert
galaxies on luminosity, based on relations from Spinoglio \etal (1995).
Spectra are plotted for luminosities between 10$^{-3}$ $L_{*}$ 
and 10$^{3}$ $L_{*}$ by factors of 10 in luminosity.

Figure 2. Predictions of diffuse IR background from galaxies.
Solid lines were calculated using the 60\mic LF;
dashed lines are from the 12\mic LF.  The upper two solid lines
show the results of our ``best estimates'' using values of $z_{flat}$
of 2 (upper solid line) and 1 (middle solid line) as described in
the text. The lower light lines are
based on our conservative lower limit assumptions about galaxy
evolution. We also show the COBE-DIRBE residuals given by Hauser (1996)
as vertical ranges,
the upper limit in the L band (log$\nu$=13.93)
given by Kashlinsky, \etal
(1996) (K),
and the upper limit band given by Stecker \& De Jager (1993) (SDJ).

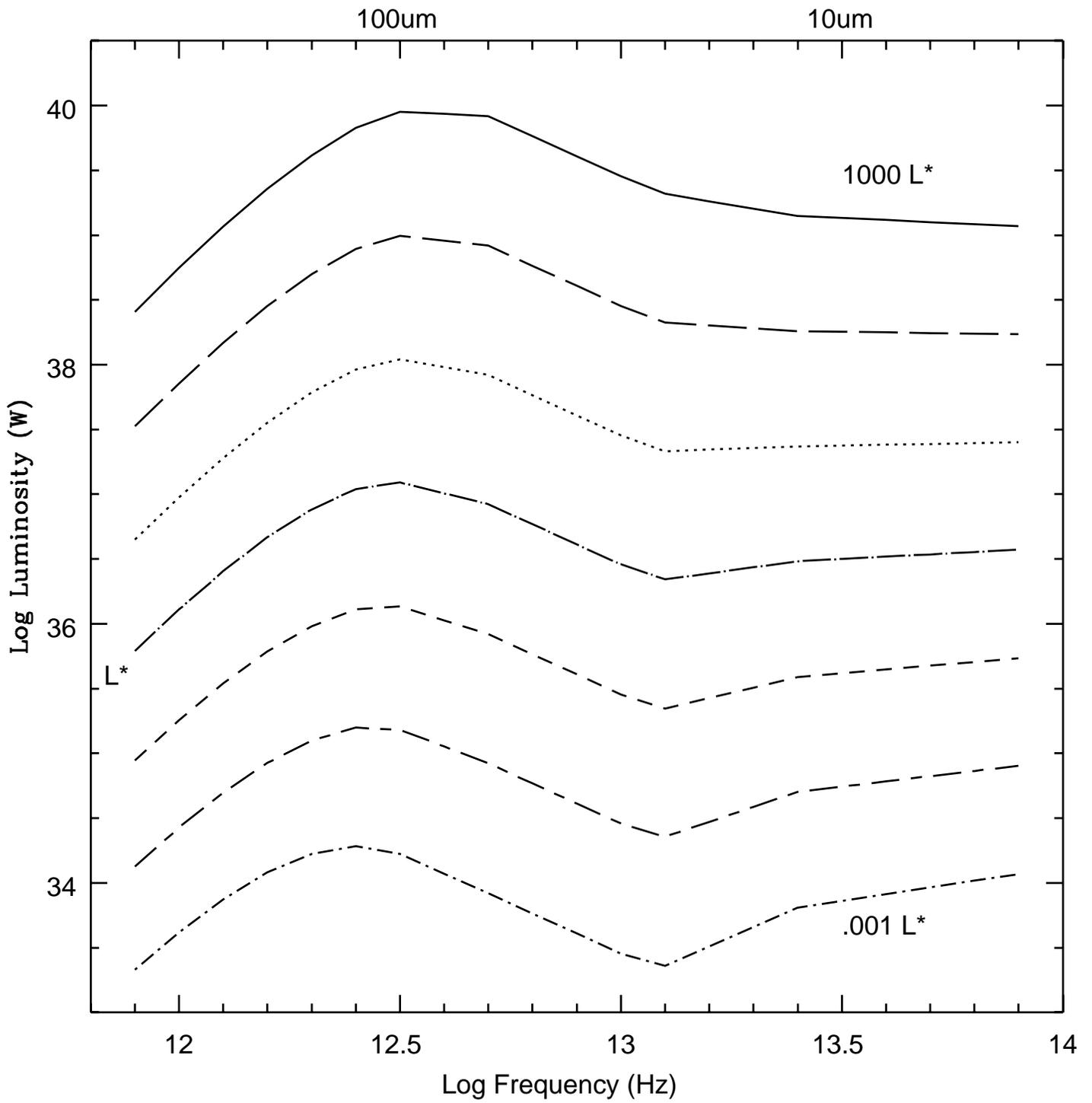
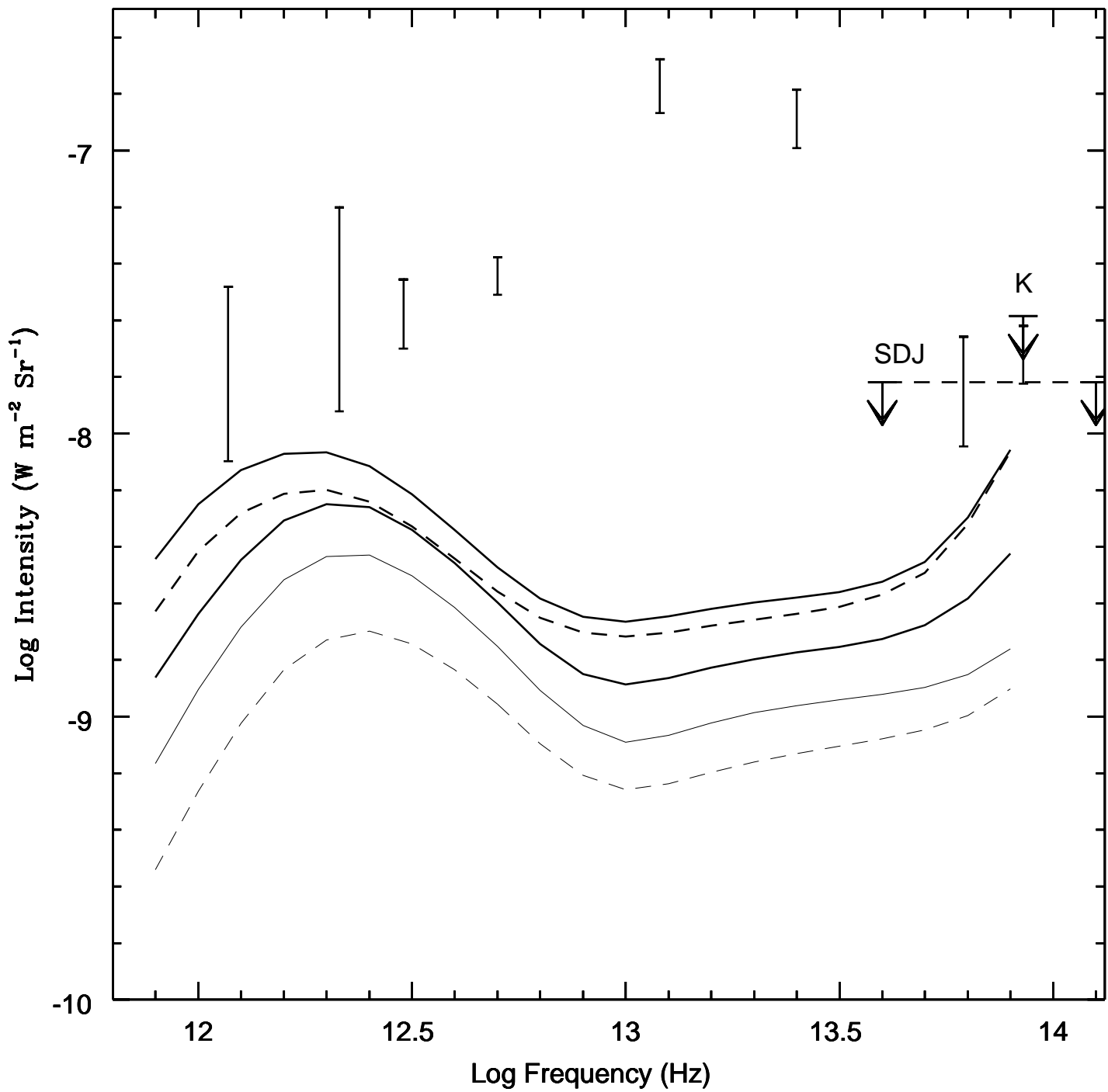
\end{document}